\documentclass[draftclsnofoot,onecolumn]{IEEEtran}
\IEEEoverridecommandlockouts

\usepackage{amssymb}
\usepackage{lettrine}
\usepackage{yhmath}

\usepackage{graphicx}
\usepackage{epstopdf}
\usepackage[normalem]{ulem}
\usepackage{balance}

\usepackage{cite}

\ifCLASSINFOpdf

\else

\fi

\usepackage{amsmath}

\usepackage{algorithm}
\usepackage{algorithmicx}
\usepackage{algpseudocode}

\usepackage{array}

\usepackage{cuted}

\usepackage{float}
\usepackage{color}
\usepackage{makecell}
\usepackage{multirow}
\usepackage{graphicx}
\usepackage{subfigure}
\usepackage{stfloats}
\usepackage{times}
\usepackage{array}
\usepackage{flushend}
\begin{document}
\pagestyle{empty}
%
\title{Active Reconfigurable Intelligent Surface Aided Surveillance Scheme }
%
%
%

\author{Xinyue Hu, Yibo Yi,\IEEEmembership{ Student Member,~IEEE,} Kun Li, Hongwei Zhang and Caihong Kai,\IEEEmembership{ Member,~IEEE}
\thanks{This work was supported by the National Natural Science Foundation of China under Grant 61971176. (Corresponding author: Caihong Kai.)}
\thanks{Xinyue Hu, Kun Li and Hongwei Zhang are with School of Electronic and Information Engineering, Anhui University, Hefei, China (e-mail: $\rm {\rm{\{ }}$21058, lik, hwzhang${\rm{\} }}$@ahu.edu.cn.).}
\thanks{ Yibo Yi and Caihong Kai are with School of Computer Science and Information Engineering, Hefei University of Technology, Hefei, China (e-mail: yiyibo2018@mail.hfut.edu.cn; chkai@hfut.edu.cn.).}

}

\maketitle


\begin{abstract}
This letter attempts to design a surveillance scheme by adopting an active reconfigurable intelligent surface (RIS). Different from the conventional passive RIS, the active RIS could not only adjust the phase
shift but also amplify the amplitude of the reflected signal. With such reflecting, the reflected signal of active RIS could jointly adjust the signal to interference plus noise ratio (SINR) of the suspicious receiver and the legitimate monitor, hence the proactive eavesdropping at
  the physical layer could be effectively realized. We formulate the optimization problem with the target of maximizing the eavesdropping rate to obtain the optimal reflecting coefficient matrix of the active RIS. {\color{blue}The formulated optimization problem is nonconvex fractional programming and challenging to deal with. We then solve the problem by approximating it as a series of convex constraints.} Simulation results validate the effectiveness of our designed surveillance scheme and show that the proposed active RIS aided surveillance scheme has good performance in terms of eavesdropping rate compared with the scheme with passive RIS.
\end{abstract}
\begin{IEEEkeywords}
Proactive eavesdropping, active reconfigurable intelligent surface, double fading, eavesdropping rate.
\end{IEEEkeywords}

%
\IEEEpeerreviewmaketitle

\section{Introduction}
\IEEEPARstart{T}{he} widespread of wireless communications not only significantly improves our life, but also brings new challenges to national security and social stability, since wireless communication links can be used by terrorists or criminals to plan and commit crimes. Thus, there is an increasing demand for authorized agencies to implement effective information surveillance to prevent crimes or terrorism attacks\cite{8014299}.

At the physical layer, the condition that the legitimate monitor could successfully eavesdrop the dubious communication is that the signal to interference plus noise ratio (SINR) at the legitimate monitor is higher than that of the suspicious destination. To achieve a nonzero eavesdropping rate \cite{8014299}, many approaches have been proposed, e.g., proactive eavesdropping via jamming\cite{9729412,9714462,9187662,9762666} and spoofing-relay based proactive eavesdropping \cite{7544447,8443394}.

 Recently, reconfigurable intelligent surface (RIS) has become a promising technique applied to secure communications \cite{9133130,Hua2022ActiveRV} and legitimate proactive eavesdropping for next-generation wireless networks \cite{9834944,9775090,9852796}. Specifically, RIS is a planar array consisting of a large number of reconfigurable passive elements, each of which can induce a certain phase shift independent of the incident signal. With such reflecting RIS can simultaneously adjust the signal to interference plus noise (SINR) of both suspicious receiver and legitimate monitor and hence enhance the eavesdropping efficiency \cite{9834944,9775090,9852796}.

{\color{blue} However, in the passive RIS aided surveillance system, the double path loss effect \cite{9306896} (signal received via the transmitter-RIS-receiver link suffer from large-scale fading twice) would lead to that the end-to-end path loss of the transmitter-RIS-suspicious destination (legitimate monitor) link is in general much larger than that of the line of sight (LoS) link from the transmitter to the suspicious destination (legitimate monitor), which is negative for using the reflection signal to ensure the successful eavesdropping condition. Moreover, to hide the legitimate monitor from the suspicious destination, the legitimate monitor should be farther from the transmitter and RIS than the suspicious destination or hidden somewhere, which will lead to that the legitimate monitor suffers more serious double path loss effect compared with the suspicious destination, i.e. the received signal by the legitimate monitor from the RIS is too weak and almost losses the function of SINR adjustment.} Recently, to overcome the double path loss effect, the active RIS technique is proposed \cite{Zhang2021ActiveRV}, in which the active reflecting element is equipped with an active power amplifier and could not only adjust the phase but also amplify the amplitude of the reflected signal. Based on the active property, the active RIS could realize significantly higher system performance, for example, channel capacity \cite{9734027}, energy efficiency \cite{9723093} and physical-layer security \cite{9652031} than those via passive RIS.

{\color{blue}Different from prior works on proactive eavesdropping, this letter proposes a novel surveillance scheme by introducing the active RIS. Since the active RIS could effectively overcome the double path loss effect, the reflected signal of the RIS is strong enough and could not only be used to ensure the successful eavesdropping condition but also jointly improve the SINR of both suspicious receiver and legitimate monitor to further increase the eavesdropping rate.} Then, we design the reflecting coefficients of the active RIS to maximize the eavesdropping rate. The formulated optimization problem is nonconvex and hard to overcome. To deal with it, we introduce a series of auxiliary variables and transfer the problem to convex formation. Simulation results show that the proposed active RIS aided surveillance scheme could obtain much higher eavesdropping rate compared with the scheme with passive RIS, which shows that our proposed scheme is an attractive option for legitimate proactive eavesdropping.

\emph{Notations}: In this letter, ${\mathop{\rm Re}\nolimits} \left( x \right)$ and ${\rm{Im}}\left( x \right)$ represent the real part and the imaginary part of $x$, respectively. $x^*$ represents the conjugate of $x$, ${{\bf{x}}^H}$ represents the conjugate transpose of the vector ${{\bf{x}}^H}$, ${{\bf{x}}^n}$ represents the $n^{th}$ element of the vector ${{\bf{x}}}$. The bandwidth term in Shannon formula is ignored, thus the unit of the channel capacity is ``bps/Hz''.

\section{System Model and Proposed F3D Aided Surveillance Scheme}
\begin{figure}[t]
  \centering
  \includegraphics[width=2.6in]{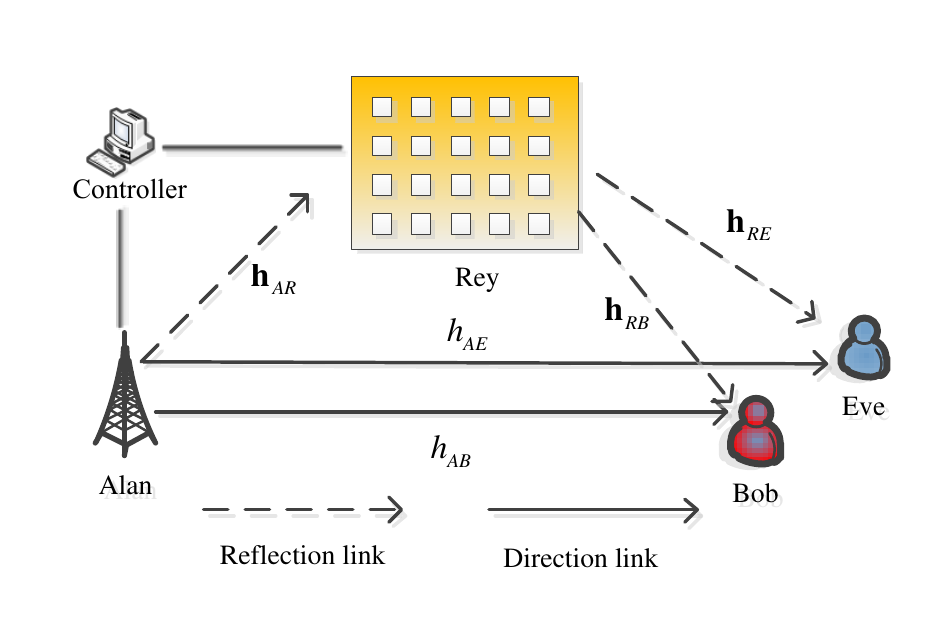}\\
  \caption{System model. }
\end{figure}
As shown in Fig. 1. We consider an active RIS aided surveillance system which consists of four nodes. A legitimate monitor Eve is intended to eavesdrop a dubious communication link from the source Alan to a suspicious destination Bob. An active RIS with $N_R$ reflecting elements, Rey, is deployed as a helper to enhance the network capacity ostensibly and secretly assist Eve to surveil the dubious communication link (i.e., the communication from Alan to Bob in this case). Except for Rey, all nodes are equipped with one antenna, and work on the half-duplex mode.

Define ${{h}_{AB}} \in\mathbb{C} $, ${h_{AE}}\in\mathbb{C}$ and ${{\bf{h}}_{AR}}\in\mathbb{C}^{N_R \times 1}$ as the channel coefficients from Alan to Bob, Eve and Rey, and those from Rey to Bob and Eve as ${{\bf{h}}_{RB}}\in\mathbb{C}^{1 \times N_R}$ and ${{\bf{h}}_{RE}}\in\mathbb{C}^{1 \times N_R}$, respectively. {\color{blue} We consider all channels are experience quasi-static block fading, and all nodes are active, hence, in each time block, via overhearing the pilot signals send by each nodes, every node could estimate the channel coefficients directly connecting with them. Moreover, Alan and Rey are both connected to a controller, Alan and Rey are going to upload channels to this controller for computing the optimal reflecting coefficients of Rey.}

Define ${s_A} \sim {\cal{CN}}(0,1)$ as the transmitted dubious information from Alan to Bob and $P_A$ as the transmit power of Alan. Rey receives $s_A$ from Alan, then reflects and amplifies the received signal, i.e., (1) to Eve and Bob, where ${\bf{\Theta }} = {\mathop{\rm diag}\nolimits} (\phi _1^{}, \ldots ,\phi _{N_R}^{})$ is the complex diagonal reflecting coefficient
matrix of Rey that adjusts the power and phase of the received signal. $\phi _n^{} = a_n^{}e_{}^{j\theta _n^{}}$ is the reflecting
coefficient of the $n^{th}$ reflecting element of Rey, where
$a_n$ and $\theta_n$ represent the amplitude and the phase, and $a_n$ can
be greater than 1 with active load. With such reflecting, the signals received by Bob and Eve are the combinations of the direct link signal from Alan and the reflected signal from Rey and can be expressed as (2) and (3), respectively.
\begin{equation}
{{\bf{y}}_R} = \sqrt {{P_A}} {{\bf{h}}_{AR}}{s_A} + {{\bf{n}}_R}.
\end{equation}

\begin{equation}
{y_B} =   \left( {{h_{AB}} + {{\bf{h}}_{RB}}{\bf{\Theta }}{h_{AR}}} \right)\sqrt {{P_A}} {s_A} + {{\bf{h}}_{RB}}{\bf{\Theta }}{{\bf{n}}_R} + {n_B}.
 \end{equation}

 \begin{equation}
{y_E} =   \left( {{h_{AE}} + {{\bf{h}}_{RE}}{\bf{\Theta }}{h_{AR}}} \right)\sqrt {{P_A}} {s_A} + {{\bf{h}}_{RE}}{\bf{\Theta }}{{\bf{n}}_R} + {n_E}.
 \end{equation}
In (1), (2) and (3), ${{\bf{n}}_R}\sim {\cal{CN}}(0,\sigma _r^2{{\bf{I}}_{{N_R}}})$ is the thermal noise generated by Rey, ${n_i} \sim {\cal{CN}}(0,\sigma _0^2),i \in \{ B,E\}$ are the AWGN at Bob and Eve, respectively.

From (1), it is easy to see that, the received power of $n^{th}$ reflecting element in Rey is $\left| {{\bf{h}}_{AR}^n} \right|_{}^2{P_A}{\rm{ + }}\sigma _r^2$ and hence the power of reflected signal from $n^{th}$ reflecting element is
 \begin{equation}
{P_n} = {\left| {{a_n}} \right|^2}\left( {\left| {{\bf{h}}_{AR}^n} \right|_{}^2{P_A}{\rm{ + }}\sigma _r^2} \right).
\end{equation}

 From (2) and (3), the signal to interference plus noise ratio (SINR) of Bob and Eve are
   \begin{equation}
\begin{array}{*{20}{l}}
{SIN{R_B} = \frac{{{P_A}{{\left| {{h_{AB}} + {{\bf{h}}_{RB}}{\bf{\Theta }}{{\bf{h}}_{AR}}} \right|}^2}}}{{\sigma _r^2{{\left| {{{\bf{h}}_{RB}}{\bf{\Theta }}} \right|}^2} + {\sigma _0^2}}}}\\
{{\rm{ = }}\frac{{{P_A}{{\bf{v}}^H}{\bf{h}}_{A - B}^H{\bf{h}}_{A - B}^{}{\bf{v}}}}{{\sigma _r^2{{\bf{v}}^H}{\rm{diag}}\left( {{{\bf{h}}_{R-B}}} \right){\rm{diag}}\left( {{{\bf{h}}_{R-B}}} \right)_{}^H{\bf{v}} + {\sigma _0^2}}}},
\end{array}
 \end{equation}
 and
  \begin{equation}
\begin{array}{*{20}{l}}
{SIN{R_E} = \frac{{{P_A}{{\left| {{h_{AE}} + {{\bf{h}}_{RE}}{\bf{\Theta }}{{\bf{h}}_{AR}}} \right|}^2}}}{{\sigma _r^2{{\left| {{{\bf{h}}_{RE}}{\bf{\Theta }}} \right|}^2} + {\sigma _0^2}}}}\\
{{\rm{ = }}\frac{{{P_A} {{{\bf{v}}^H}{\bf{h}}_{A - E}^H{\bf{h}}_{A - E}^{}{\bf{v}}} }}{{\sigma _r^2{{\bf{v}}^H}{\rm{diag}}\left( {{{\bf{h}}_{R-E}}} \right){\rm{diag}}\left( {{{\bf{h}}_{R-E}}} \right)_{}^H{\bf{v}} + {\sigma _0^2}}}},
\end{array}
  \end{equation}
respectively, where ${\bf{h}}_{A - B}^{} = \left[ {{{\bf{h}}_{RB}}{\rm{diag}}\left( {{{\bf{h}}_{AR}}} \right),{h_{AB}}} \right]$, ${\bf{h}}_{A - E}^{} = \left[ {{{\bf{h}}_{RE}}{\rm{diag}}\left( {{{\bf{h}}_{AR}}} \right),{h_{AE}}} \right]$, ${{\bf{h}}_{R - B}} = \left[ {{{\bf{h}}_{RB}},0} \right]$, ${{\bf{h}}_{R - E}} = \left[ {{{\bf{h}}_{RE}},0} \right]$ and ${\bf{v}} = \left[ {\phi _1^{}, \ldots ,\phi _{N_R}^{},1} \right]_{}^T$

\section{Optimization of the RIS Aided Surveillance Scheme}

We next formulate our eavesdropping rate maximization problem and design algorithms to solve it. In particular, we make attempts to find the optimal reflecting coefficient $\bf{v}$ to maximize the eavesdropping rate.
\subsection{Problem Formulation}
We follow two principles for designing the optimization problem of the proposed surveillance scheme: 1). Guaranteeing Eve to successfully eavesdrop the communication of Bob. 2). Maximizing the eavesdropping rate. The first principle requires that $SIN{R_E}\ge SIN{R_B}$ for ensuring Eve can receive and decode the dubious information at the physical layer. The second principle represents that $SIN{R_B}$ should be as high as possible when $SIN{R_E}\ge SIN{R_B}$ is satisfied\footnote{The eavesdropping rate is $\log (1 + SIN{R_B})$ when $SIN{R_E}\ge SIN{R_E}$, thus the larger the $SIN{R_B}$, the higher the eavesdropping rate. Also, since Rey is deployed to enhance the network capacity ostensibly in the considered scenario, the $SIN{R_B}$ should be increased under the proposed scheme, otherwise, Bob might suspect he is being bugged.}.

Based on the two principles, the following optimization problem is formulated to obtain the optimal reflecting coefficient $\bf{v}$\footnote{Due to the monotonicity of the logarithmic function, we omit the $\rm log_2$ in the formulated promlem}:
 \begin{equation}
\begin{array}{*{20}{l}}
{{\cal P}1:\mathop {\max }\limits_{\bf{v}} SIN{R_B}}\\
\begin{array}{l}
{\rm{s}}.{\rm{t}}.{\rm{C1}}:{\left| {{{\bf{v}}^n}} \right|^2} \le \frac{{{P_{\max }}}}{{\left( {\left| {{\bf{h}}_{AR}^n} \right|_{}^2{P_A}{\rm{ + }}\sigma _r^2} \right)}},n = 1, \ldots ,N_R\\
\;\;\;\;\;{\rm{C2}}:{\rm{ }}{{\bf{v}}^{N + 1}} = 1
\end{array}\\
{\;\;\;\;\;\;{\rm{C3}}:SIN{R_E} \ge SIN{R_B}},
\end{array}
\end{equation}
where $P_{\rm max}$ is the maximal amplification power budget of each reflecting element in Rey. In ${\cal P}1$, the objective is to maximize the eavesdropping rate under the reflecting power constraint of each element in Rey, i.e., C1, while making sure that Eve could successfully eavesdrop the communication of Bob, i.e., C3.
\subsection{Optimal Solution of P1}
In ${\cal P}1$, we can see that C1 and C2 are convex constraints, {\color{blue} however, different from the passive RIS schemes, due to the active property of Rey, the introduced additional thermal noise ${\bf n}_R$ creates additional interference at Bob and Eve, and thus the objective function and constraint C3 of ${\cal P}1$ become nonconvex quadratic fractional form ($\bf{v}$  appears on the both numerator and denominator of (5) and (6)) and complicated.}

{\color{blue} To deal with ${\cal P}1$, we first introduce a group of auxiliary variable ${a,b}$ to convert the nonconvex objective function to a linear form and new auxiliary constraints, i.e.,}

 \begin{equation}
\begin{array}{*{20}{l}}
{{\cal P}2:\mathop {\max }\limits_{{\bf{v}},a,b} a}\\
{\begin{array}{*{20}{l}}
{{\rm{s}}.{\rm{t}}.{\rm{C1}},{\rm{C2}},{\rm{C3}}}\\
{\;\;\;\;\;\rm{C4}}:SIN{R_B} \ge a \Leftrightarrow \frac{{{P_A}{{\bf{v}}^H}{\bf{h}}_{A - B}^H{\bf{h}}_{A - B}^{}{\bf{v}}}}{b} \ge a\\
{\;\;\;\;\;{\rm{C}}5:b \ge \sigma _r^2{{\bf{v}}^H}{\rm{diag}}\left( {{{\bf{h}}_{R-B}}} \right){\rm{diag}}\left( {{{\bf{h}}_{R-B}}} \right)_{}^H{\bf{v}} + {n_0}}.\\
\end{array}}\\
\end{array}
\end{equation}
{\color{blue} Now, with the aided of constraints C4 and C5, the objective function is equivalently converted to linear.}

It is easy to see that C5 can be directly transformed to second-order cone (SOC) form as
 \begin{equation}
{\rm{C}}5':\frac{{b - {n_0} + 1}}{2} \ge \left\| {\left[ {\left| {\sigma _r^{}{\rm{diag}}\left( {{{\bf{h}}_{R-B}}} \right)_{}^H{\bf{v}}} \right|,\frac{{b - {n_0} - 1}}{2}} \right]} \right\|_2^{}.
\end{equation}

Although C4 is nonconvex, fortunately, it is in the form of {\color{blue}${{x^2}/y} \ge a $} which is convex, hence by performing first-order Taylor series expansion, C4 can be approximated to the following linear constraint:
 \begin{equation}
{\rm{C4'}}:2{\mathop{\rm Re}\nolimits} \left( {\frac{{{P_A}{\bf{v}}_0^H{\bf{H}}_{A - B}^{}}}{{b_0^{}}}{\bf{v}}} \right) - \frac{{{P_A}{\bf{v}}_0^H{\bf{H}}_{A - B}^{}{\bf{v}}_0^{}}}{{b_0^2}}b \ge a,
\end{equation}
where ${\bf{H}}_{A - B}^{} = {\bf{h}}_{A - B}^H{\bf{h}}_{A - B}^{}$.

With the aid of ${\rm{C4'}}$ and ${\rm{C5'}}$, the nonconvex objective function of the primal ${\cal P}1$ becomes linear. Next, we transfer the nonconvex C3 to a series of new constraints by introducing another auxiliary variable $c$, i.e., ${\rm{C}}6:SIN{R_E} \ge c$ and ${\rm{C}}7:c \ge SIN{R_B}$. {\color{blue}Although ${\rm{C}}6$ and ${\rm{C}}7$ are still nonconvex, they be becomes simpler compared with C3.}

 Note that, ${\rm{C}}6$ and C4 have the same form, hence with the similar approach that we deal with C4, ${\rm{C}}6$ could be approximated to C8 and C9 which have the same form with ${\rm{C4'}}$ and ${\rm{C5'}}$, respectively. For briefly we ignore C8 and C9 here.

For overcoming nonconvex fractional constraint C7, we introduce a new auxiliary variable $d$ and then rewrite C7 to

 \begin{equation}
{\rm{C}}10:cd \ge {\left| {\sqrt {{P_A}} {\bf{h}}_{A - B}^{}{\bf{v}}} \right|^2},
\end{equation}
and
 \begin{equation}
{\rm{C}}11:\sigma _r^2{{\bf{v}}^H}{\rm{diag}}\left( {{{\bf{h}}_{R - B}}} \right){\rm{diag}}\left( {{{\bf{h}}_{R - B}}} \right)_{}^H{\bf{v}} + {n_0} \ge d,
\end{equation}
respectively. Further C10 can be directly transformed to SOC form as
 \begin{equation}
{\rm{C}}10':\frac{{c + d}}{2} \ge {\left\| {\left[ {\frac{{c - d}}{2},\left| {\sqrt {{P_A}} {\bf{h}}_{A - B}^{}{\bf{v}}} \right|} \right]} \right\|_2}.
\end{equation}
Then, by performing the first-order Taylor series expansion, C11 can be approximated to
 \begin{equation}
{\rm{C}}11':\sigma _r^22{\mathop{\rm Re}\nolimits} \left( {{\bf{v}}_0^H{{\bf{H}}_{R - B}}{{\bf{v}}^{}}} \right) - \sigma _r^2{\bf{v}}_0^H{{\bf{H}}_{R - B}}{\bf{v}}_0 + {n_0} \ge d,
\end{equation}
where ${{\bf{H}}_{R - B}}{\rm{ = diag}}\left( {{{\bf{h}}_{R - B}}} \right){\rm{diag}}\left( {{{\bf{h}}_{R - B}}} \right)_{}^H$.

Now, all non-convex constraints have been approximated to convex constraints, the approximated convex version of ${\cal P}1$ is given by
\begin{equation}
\begin{array}{*{20}{l}}
{{\cal P}3:\mathop {\max }\limits_{{\bf{v}},a,b,c,d} a}\\
{{\rm{s}}.{\rm{t}}.{\rm{C1,C2,C4',C5',C8,C9,C10',C11'}}},
\end{array}
\end{equation}
which can be solved by using interior point method. Finally, we can solve the original ${\cal P}1$ by iterating ${\bf{v}}_0^{}$ in ${\cal P}3$.

{\color{blue}{\it Complexity: } The complexity of interior point method for solving the convex optimization problem ${\cal P}3$ is $O\left( {{\sigma ^{0.5}}(\sigma  + \varsigma ){\varsigma ^2}} \right)$, in which $\sigma {\rm{ = }}{\rm{7}}$ is the number of inequality constraints and $\varsigma={2N_R+5}$ is the number of optimization variables. Combining with the fact that the iterative number is bounded by $R_{}^{\max }$, we have the total computational complexity for solving ${\cal P}1$ is $O\left( R_{}^{\max }{{\sigma ^{0.5}}(\sigma  + \varsigma ){\varsigma ^2}} \right)$.}
\subsection{A Time-Saving Suboptimal Approach to Solve ${\cal P}\text{1}$}

In Section \uppercase\expandafter{\romannumeral3}-B, for obtaining the optimal reflecting coefficient matrix ${\bf{\Theta }} = {\mathop{\rm diag}\nolimits} (\phi _1^{}, \ldots ,\phi _{N_R.})$, interior point method is used to solve the convex problem ${\cal P}3$, meanwhile ${\bf{v}}_0^{}$ should be iterated several times, which may be time-consuming. In this subsection, by simplifying ${\cal P}\text{1}$ properly, we can directly obtain a suboptimal $\phi _n$ in ${\bf{\Theta }}$ by fixing other element in ${\bf{\Theta }}$, and then the suboptimal solution of ${\cal P}\text{1}$ can be achieved by respectively solving each $\phi _n$ in ${\bf{\Theta }}$.

Let us focus on $\phi _n$, then (5) and (6) can be rewritten to (16) and (17) on the bottom of the next page, respectively, where ${\phi _{n,r}}$ and ${\phi _{n,i}}$ are the real part and imaginary part of $\phi _n$, respectively, ${h_{AB,\backslash n}} = {\left| {{h_{AB}} + \sum\limits_{j \ne n}^{} {{\bf{h}}_{AR}^j{\bf{h}}_{RB}^j\phi _j^{}} } \right|^2}$, ${h_{RB,\backslash n}} = \sum\limits_{j \ne n}^{} {\left| {{\bf{h}}_{RB}^j\phi _j^{}} \right|_{}^2} $, ${h_{AE,\backslash n}} = {\left| {{h_{AE}} + \sum\limits_{j \ne n}^{} {{\bf{h}}_{AR}^j{\bf{h}}_{RE}^j\phi _j^{}} } \right|^2}$ and ${h_{RE,\backslash n}} = \sum\limits_{j \ne n}^{} {\left| {{\bf{h}}_{RE}^j\phi _j^{}} \right|_{}^2} $.
\begin{figure*}[b]
\vspace*{2pt}
\hrulefill 

\begin{equation*}
SIN{R_B}\left(\! {\phi _{n,r}^{},\!\phi _{n,i}^{}} \!\right)\! =\! 1{\rm{\! +\! }}\frac{{{P_A}\left(\! {{{\left|\! {{\bf{h}}_{AR}^n{\bf{h}}_{RB}^n} \!\right|}^2}\left( \!{\phi _{n,r}^2 \!+ \!\phi _{n,i}^2} \!\right)\! +\! {h_{AB,\backslash n}}\! +\! 2{\rm{Re}}\left( \! {{\bf{h}}_{AR}^n{\bf{h}}_{RB}^nh_{AB,\backslash n}^*}\! \right)\phi _{n,r}^{} \!-\! 2{\rm{Im}}\left( {{\bf{h}}_{AR}^n{\bf{h}}_{RB}^nh_{AB,\backslash n}^*} \right)\phi _{n,i}^{}} \right)}}{{{{\sigma _r^2}{\left| {{\bf{h}}_{RB}^n} \right|}^2}\left( {\phi _{n,r}^2 \!+ \!\phi _{n,i}^2} \right) + {\sigma _r^2}{h_{RB,\backslash n}} + {{\sigma _0^2}}}}\tag{16}
\end{equation*}

\end{figure*}
\begin{figure*}[b]
\hrulefill 

\begin{equation*}
SIN{R_E}\left(\! {\phi _{n,r}^{},\!\phi _{n,i}^{}}\! \right)\! =\! 1{\rm{ \!+\! }}\frac{{{P_A}\left( \!{{{\left|\! {{\bf{h}}_{AR}^n{\bf{h}}_{RE}^n} \!\right|}^2}\left( \!{\phi _{n,r}^2 \!+\! \phi _{n,i}^2} \!\right)\! +\! {h_{AE,\backslash n}}\! +\! 2{\rm{Re}}\left(\! {{\bf{h}}_{AR}^n{\bf{h}}_{RE}^nh_{AE,\backslash n}^*}\! \right)\phi _{n,r}^{}\! -\! 2{\rm{Im}}\left( \!{{\bf{h}}_{AR}^n{\bf{h}}_{RE}^nh_{AE,\backslash n}^*} \right)\phi _{n,i}^{}}\! \right)}}{{\sigma _r^2{{\left| {{\bf{h}}_{RE}^n} \right|}^2}\left( {\phi _{n,r}^2 + \phi _{n,i}^2} \right) + \sigma _r^2{h_{RE,\backslash n}} + \sigma _0^2}}\tag{17}
\end{equation*}

\end{figure*}
Considering the worst case of the surveillance system, in which Eve suffers the maximal interference caused by the thermal noise generated
by the $n^{th}$ element of Rey, while Bob does not suffer that noise. In this case ${\cal P}\text{1}$ becomes
\begin{equation}
 \setcounter{equation}{18}
\begin{array}{*{20}{l}}
{{\cal P}4:\mathop {\max }\limits_{{\phi _{n,r}},{\phi _{n,i}}} \overline {SIN{R_B}} \left( {{\phi _{n,r}},{\phi _{n,i}}} \right)}\\
{{\rm{s}}.{\rm{t}}.:\overline {{\rm{C1}}} :\phi _{n,r}^2 + \phi _{n,i}^2 \le \frac{{{P_{\max }}}}{{\left( {\left| {{\bf{h}}_{AR}^n} \right|_{}^2{P_A}{\rm{ + }}\sigma _r^2} \right)}}}\\
{\;\;\;\;\;\;\;\;\overline {{\rm{C3}}} :\overline {SIN{R_E}} \left( {{\phi _{n,r}},{\phi _{n,i}}} \right) \ge \overline {SIN{R_B}} \left( {{\phi _{n,r}},{\phi _{n,i}}} \right)},
\end{array}
\end{equation}
in which ${\overline {SIN{R_E}} \left( {\phi _{n,r}},{\phi _{n,i}} \right)}$ is obtained by setting ${\phi _{n,r}^2 + \phi _{n,i}^2}$ in the denominator of (16) as ${P_{\max }}/\left( {\left| {{\bf{h}}_{AR}^n} \right|_{}^2{P_A}{\rm{ + }}\sigma _r^2} \right)$ (equals to Eve suffers the maximal interference caused by the thermal noise generated by the $n^{th}$ element of Rey), ${\overline {SIN{R_B}} \left( {\phi _{n,r}},{\phi _{n,i}}  \right)}$ is obtained by setting ${\phi _{n,r}^2 + \phi _{n,i}^2}$ in the denominator of (17) as $0$ (equals to Bob does not suffer the thermal noise of $n^{th}$ element in Rey). Since ${\overline {{\rm{C3}}} }$ is more stringent than ${{\rm{C3}}}$ in ${\cal P}\text{1}$, the optimal solution of ${\cal P}\text{4}$ must be a feasible solution of ${\cal P}\text{1}$. In the following, we try to solve the optimal solution of ${\cal P}\text{4}$ and then find out a suboptimal ${\bf{\Theta }}$.

\begin{figure}[t]
  \centering
  \includegraphics[width=2.8in]{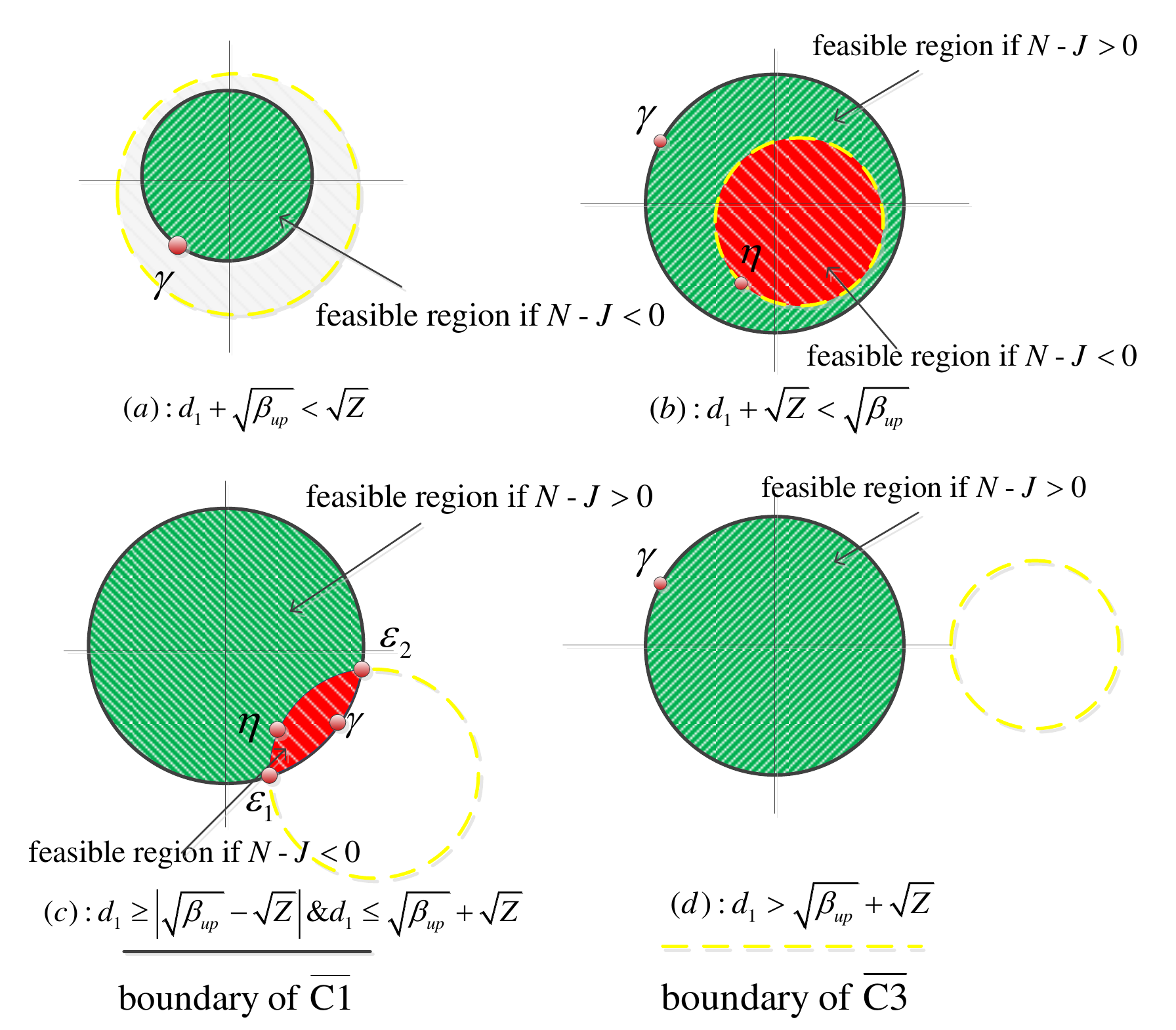}\\
  \caption{The feasible region and the candidate solution points of ${\cal P}5$. }
\end{figure}
By introducing auxiliary variables $J, K, L, M, N, O, P$, $Q$ and ignoring the case that $N-J=0$, ${\cal P}\text{4}$ could be simplified to
 \begin{equation}
\begin{array}{*{20}{l}}
{\cal P}{\rm{5}}:\mathop {\max }\limits_{{\phi _{n,r}},{\phi _{n,i}}} \overline {SIN{R_B}} \! = \!J\left(\! {\phi _{n,r}^2\! +\! \phi _{n,i}^2}\! \right)\! +\! K \!+\! L{\phi _{n,r}}\! + \!M{\phi _{n,i}}\\
{{\rm{s}}.{\rm{t}}.:\overline {{\rm{C1}}} }\\
{\;\;\;\;\;\;\;\;\overline {{\rm{C3}}} :\left(\! {N \!-\! J}\! \right)\left( \!{\phi _{n,r}^2{\rm{\! +\! }}\phi _{n,i}^2} \!\right)\! + \!\left(\! {P\! -\! L}\! \right){\phi _{n,r}}\! +\! \left( \!{Q\! -\! M}\! \right){\phi _{n,i}}}\\
{\;\;\;\;\;\;\; \ge K - O}\\
{\;\; \Rightarrow \left\{ {\begin{array}{*{20}{l}}
{\overline {{\rm{C31}}} :{{\left( {{\phi _{n,r}} -  - \frac{{\left( {P - L} \right)}}{{2\left( {N - J} \right)}}} \right)}^2} + {{\left( {{\phi _{n,i}} -  - \frac{{\left( {Q - M} \right)}}{{2\left( {N - J} \right)}}} \right)}^2}}\\
{ \ge \frac{{K - O}}{{\left( {N - J} \right)}} + \frac{{\left( {P - L} \right)_{}^2}}{{4\left( {N - J} \right)_{}^2}} + \frac{{\left( {Q - M} \right)_{}^2}}{{4\left( {N - J} \right)_{}^2}},{\rm{if}}\;\;N - J > 0}\\
{\overline {{\rm{C32}}} :{{\left( {{\phi _{n,r}} -  - \frac{{\left( {P - L} \right)}}{{2\left( {N - J} \right)}}} \right)}^2} + {{\left( {{\phi _{n,i}} -  - \frac{{\left( {Q - M} \right)}}{{2\left( {N - J} \right)}}} \right)}^2}}\\
{ \le \frac{{K - O}}{{\left( {N - J} \right)}} + \frac{{\left( {P - L} \right)_{}^2}}{{4\left( {N - J} \right)_{}^2}} + \frac{{\left( {Q - M} \right)_{}^2}}{{4\left( {N - J} \right)_{}^2}},{\rm{if}}\;\;N - J < 0},
\end{array}} \right.}
\end{array}
\end{equation}
in which the auxiliary variables are represented in (20) on the bottom of the next page.
\begin{figure*}[b]

\hrulefill 
\begin{small}
\begin{equation*}
\begin{array}{*{20}{l}}
{J = \frac{{{P_A}{{\left| {{\bf{h}}_{AR}^n{\bf{h}}_{RB}^n} \right|}^2}}}{{\sigma _r^2{h_{RB,\backslash n}} + \sigma _0^2}},K = \frac{{{P_A}{h_{AB,\backslash n}}}}{{\sigma _r^2{h_{RB,\backslash n}} + \sigma _0^2}},L = \frac{{{P_A}2{\rm{Re}}\left( {{\bf{h}}_{AR}^n{\bf{h}}_{RB}^nh_{AB,\backslash n}^*} \right)}}{{\sigma _r^2{h_{RB,\backslash n}} + \sigma _0^2}},M = \frac{{ - 2{P_A}{\rm{Im}}\left( {{\bf{h}}_{AR}^n{\bf{h}}_{RB}^nh_{AB,\backslash n}^*} \right)}}{{\sigma _r^2{h_{RB,\backslash n}} + \sigma _0^2}},N = \frac{{{P_A}{{\left| {{\bf{h}}_{AR}^n{\bf{h}}_{RE}^n} \right|}^2}}}{{\sigma _r^2{{\left| {{\bf{h}}_{RE}^n} \right|}^2}{\beta _{up}} + \sigma _r^2{h_{RE,\backslash n}} + \sigma _0^2}}}\\
{O = \frac{{{P_A}{h_{AE,\backslash n}}}}{{\sigma _r^2{{\left| {{\bf{h}}_{RE}^n} \right|}^2}{\beta _{up}} + \sigma _r^2{h_{RE,\backslash n}} + \sigma _0^2}},P = \frac{{{P_A}2{\rm{Re}}\left( {{\bf{h}}_{AR}^n{\bf{h}}_{RE}^nh_{AE,\backslash n}^*} \right)}}{{\sigma _r^2{{\left| {{\bf{h}}_{RE}^n} \right|}^2}{\beta _{up}} + \sigma _r^2{h_{RE,\backslash n}} + \sigma _0^2}},Q = \frac{{ - 2{P_A}{\rm{Im}}\left( {{\bf{h}}_{AR}^n{\bf{h}}_{RE}^nh_{AE,\backslash n}^*} \right)}}{{\sigma _r^2{{\left| {{\bf{h}}_{RE}^n} \right|}^2}{\beta _{up}} + \sigma _r^2{h_{RE,\backslash n}} + \sigma _0^2}}}
\end{array}\tag{20}
\end{equation*}
\end{small}
\end{figure*}

For solving ${\cal P}\text{5}$, we first rewrite the objective function as
 \begin{equation}
  \setcounter{equation}{21}
{({\phi _{n,i}}^{} \!- \! -\! \frac{L}{{2J}})^2}\! +\! {({\phi _{n,r}}^{} \!- \! -\! \frac{M}{{2J}})^2}\! =\! \frac{\overline{{SIN{R_B}}} - K}{J} \!+\! \frac{{L_{}^2}}{{4J_{}^2}}\! +\! \frac{{M_{}^2}}{{4J_{}^2}},
\end{equation}
which means the point set of $({\phi _{n,i}},{\phi _{n,r}})$ that make the object function equals to $ \overline{SIN{R_B}}$ is a ring with central $\left( { - \frac{L}{{2J}}, - \frac{M}{{2J}}} \right) = (S,T)$ and square of radius ${\frac{{\overline{SIN{R_B}} - K}}{J} + \frac{{L_{}^2}}{{4J_{}^2}} + \frac{{M_{}^2}}{{4J_{}^2}}} $. Since the square of radius above is monotone increasing with $\overline{SIN{R_B}}$, we have the optimal solution of ${\cal P}\text{5}$ is a point $({\phi _{n,i}},{\phi _{n,r}})$ which is furthest from $(S,T)$ in the feasible region $\overline {{\rm{C1}}}  \cap \overline {{\rm{C}}3}$.

We next check out the geometrical features of $\overline {{\rm{C1}}}  \cap \overline {{\rm{C}}3}$ for finding the optimal solution of ${\cal P}\text{5}$. It is easy to see that the  boundary of $\overline {{\rm{C}}2}$ is a ring with central $\left( { - \frac{{\left( {P - L} \right)}}{{2\left( {{{N - J}}} \right)}}, - \frac{{\left( {Q - M} \right)}}{{2\left( {{{N - J}}} \right)}}} \right) = (U,W)$ and square of radius ${\frac{{K - O}}{{\left( {N - J} \right)}} + \frac{{\left( {P - L} \right)_{}^2}}{{4\left( {N - J} \right)_{}^2}} + \frac{{\left( {Q - M} \right)_{}^2}}{{4\left( {N - J} \right)_{}^2}}}  = Z$. When $N-J>0$, the feasible region of $\overline {{\rm{C}}3}$ is the outer region of the above ring (i.e., $\overline {{\rm{C}}31}$ ), vice versa (i.e., $\overline {{\rm{C}}32}$).
Let $\gamma$ be the point furthest from $(S,T)$ in the circle $\overline {{\rm{C}}1}$, $\eta $ be the point furthest from $(S,T)$ in the circle $\overline {{\rm{C}}3}$ (if it exists), ${\varepsilon _1} $ and ${\varepsilon _2}$ be the intersection points of the boundary of $\overline {{\rm{C}}1}$ with $\overline {{\rm{C}}3}$ (if it exists). In the following we investigate the geometrical features of $\overline {{\rm{C1}}}  \cap \overline {{\rm{C}}3}$ case by case and give the optimal solution ${\phi _{n}^{}}$ of ${\cal P}\text{5}$ in each case.
\subsubsection{$N-J>0$ and $Z\le0$}
In this case, $\overline{{\rm{C}}31}$ must be the entire complex plane, since the left hand side of $\overline {{\rm{C}}31}$ always greater than or equal to 0, which means $\overline {{\rm{C1}}}  \cap \overline {{\rm{C}}3}$ is equal to  $\overline {{\rm{C}}1}$ and ${\phi _{n}^{}}$ is $\gamma$.
\subsubsection{$N-J>0$ and $Z>0$}
Let $d_1$ be the distance between $(U,W)$ and $(0,0)$, ${\beta _{up}} = {P_{\max }}/\left( {\left| {{\bf{h}}_{AR}^n} \right|_{}^2{P_A}{\rm{ + }}\sigma _r^2} \right)$, in this case, the feasible region could be deposed into four subcases:
 \begin{itemize}
\item[\bf{1}.] If ${d_1}+\sqrt{{\beta _{up}}}<\sqrt {Z}$, as shown in Fig. 2(a) $\overline {{\rm{C1}}}$ falls into the unfeasible region of $\overline {{\rm{C31}}}$, thus the solution of ${\cal P}\text{5}$ does not exist and we set ${\phi _{n}^{}}=(0,0)$.
\item[\bf{2}.] As in Fig. 2(b), if ${d_1}+\sqrt {Z}<\sqrt{{\beta _{up}}}$, the whole boundary of $\overline {{\rm{C1}}}$ belongs to the feasible region, hence, ${\phi _{n}^{}}$ is $\gamma$.
\item[\bf{3}.]  As in Fig. 2(d), if ${d_1}>\sqrt{{\beta _{up}}}+\sqrt {Z}$, the feasible region $\overline {{\rm{C1}}}  \cap \overline {{\rm{C}}3}$ is equal to $\overline {{\rm{C}}1}$, which means ${\phi _{n}^{}}$ is $\gamma$.
\item[\bf{4}.] As in Fig. 2(c), if ${d_1}\! \ge \!\left| {\sqrt {{\beta _{up}}} \! -\! \sqrt {Z}} \right|\& d_1 \!\le\! \sqrt {{\beta _{up}}} \! +\! \sqrt {Z}$, $\overline {{\rm{C1}}}  \cap \overline {{\rm{C}}3}$ is the part of $\overline {{\rm{C1}}}$ that is not obscured by the unfeasible region of $\overline {{\rm{C31}}}$. In this case ${\phi _{n}^{}}$ must be one of $\gamma$, ${\varepsilon _1} $ and ${\varepsilon _2}$, in detail, if $\gamma$ satisfies $\overline {{\rm{C31}}}$, then ${\phi _{n}^{}}=\gamma$, otherwise, ${\phi _{n}^{}}$ is one of ${\varepsilon _1} $ and ${\varepsilon _2}$ that is closer to $\gamma$.

 \end{itemize}
\subsubsection{$N-J<0$ and $Z\le0$}
In this case, $\overline {{\rm{C32}}}$ must be null set or (0,0), since the left hand side of $\overline {{\rm{C32}}}$ must be greater than or equal to zero and hence $\overline {{\rm{C1}}}  \cap \overline {{\rm{C}}3}$ is null set or (0,0). Thus in this case the solution of ${\cal P}\text{5}$ does not exist or equals to (0,0) and we set ${\phi _{n}^{}}=(0,0)$.

\subsubsection{$N-J<0$ and $Z>0$}
In this case, the feasible region also could be deposed into four subcases:
 \begin{itemize}
\item[\bf{1}.] As in Fig. 2(a), if ${d_1}+\sqrt{{\beta _{up}}}<\sqrt {Z}$, $\overline {{\rm{C1}}}$ falls into the feasible region of $\overline {{\rm{C32}}}$, thus $\overline {{\rm{C1}}}$ is a subset of $\overline {{\rm{C31}}}$ i.e., $\overline {{\rm{C1}}}  \cap \overline {{\rm{C}}3}$=$\overline {{\rm{C1}}}$ and ${\phi _{n}^{}}$ is $\gamma$.

\item[\bf{2}.] As in Fig. 2(b), if ${d_1}+\sqrt {Z}<\sqrt{{\beta _{up}}}$, $\overline {{\rm{C32}}}$ falls into the feasible region of $\overline {{\rm{C1}}}$, thus $\overline {{\rm{C32}}}$ is a subset of $\overline {{\rm{C1}}}$ i.e., $\overline {{\rm{C1}}}  \cap \overline {{\rm{C}}2}$=$\overline {{\rm{C32}}}$ and ${\phi _{n}^{}}$ is $\eta$.
\item[\bf{3}.] If ${d_1}>\sqrt{{\beta _{up}}}+\sqrt {Z}$, the feasible region $\overline {{\rm{C1}}}  \cap \overline {{\rm{C}}3}$ is null set which means ${\cal P}\text{5}$ does not exist and we set ${\phi _{n}^{}}=(0,0)$.
\item[\bf{4}.] As in Fig. 2(c), if ${d_1}\! \ge \!\left| {\sqrt {{\beta _{up}}} \! -\!\sqrt {Z}} \right|\& d_1 \!\le\! \sqrt {{\beta _{up}}} \! +\! \sqrt {Z}$,  $\overline {{\rm{C1}}}  \cap \overline {{\rm{C}}3}$ is the overlap of $\overline {{\rm{C1}}}$ and  $\overline {{\rm{C}}32}$. In this case ${\phi _{n}^{}}$ is one of $\gamma$, $\eta$, ${\varepsilon _1} $ and ${\varepsilon _2}$, in detail, if $\gamma$ satisfies $\overline {{\rm{C32}}}$, then ${\phi _{n}^{}}=\gamma$, else if $\eta$ satisfies $\overline {{\rm{C1}}}$, then ${\phi _{n}^{}}=\eta$, otherwise ${\phi _{n}^{}}$ is one of ${\varepsilon _1} $ and ${\varepsilon _2}$ that makes $\overline {SINR_B}$ higher.
 \end{itemize}

Up to now, by fixing $\phi _j^{},j \ne n$, we could directly obtain the suboptimal $\phi _n^{}$. By iteratively solving each ${\phi _{n}^{}}$ with several rounds the suboptimal ${\bf{\Theta }}$ could be obtained. The procedure for solving ${\cal P}5$ is summarized in Algorithm 1.

{\color{blue}{\it Complexity:} Only one time calculation is enough to obtain the suboptimal solution of ${\phi _{n}^{}}$, thus, the total computational complexity of the suboptimal algorithm is $O\left( R_{}^{\max }N_R\right)$, where $ R_{}^{max}$ is the maximum iteration number.}

\begin{algorithm}
\caption{Solving ${\cal P}4$}

\label{alg1}
\begin{algorithmic}[1]
\State \textbf{Initialization:} Calculating $d_1$, $\beta_{up}$, $\gamma$, $\eta $, ${\varepsilon _1} $ and ${\varepsilon _2} $ based on the geometrical features of $\overline {{\rm{C1}}}  \cap \overline {{\rm{C}}3}$.

  \If{$N-J>0$ and $Z\le0$}
     \State  ${\phi _{n}^{}}=\gamma$.
  \ElsIf {$N-J>0$ and $Z>0$}
    \If{${d_1} + \sqrt {{\beta _{up}}}  < \sqrt Z $}
      \State ${\phi _{n}^{}}=(0,0)$.
    \ElsIf {${d_1} + \sqrt Z  < \sqrt {{\beta _{up}}}  $}
      \State ${\phi _{n}^{}}=\gamma$.
    \ElsIf {${d_1} >\sqrt Z+\sqrt {{\beta _{up}}}  $}
      \State ${\phi _{n}^{}}=\gamma$.
     \ElsIf {${d_1} \ge \left| {\sqrt {{\beta _{up}}}  - \sqrt Z } \right|\& {d_1} \le \sqrt {{\beta _{up}}}  + \sqrt Z   $}
      \State If $\gamma$ satisfies $\overline {{\rm{C31}}}$, then ${\phi _{n}^{}}=\gamma$, otherwise, ${\phi _{n}^{}}$ is
      \State one of ${\varepsilon _1} $ and ${\varepsilon _2}$ that is closer to $\gamma$.
      \EndIf
   \ElsIf {$N-J<0$ and $Z\le0$}
       \State  ${\phi _{n}^{}}=(0,0)$
   \ElsIf {$N-J<0$ and $Z>0$}
    \If{${d_1} + \sqrt {{\beta _{up}}}  < \sqrt Z $}
      \State ${\phi _{n}^{}}=\gamma$.
    \ElsIf {${d_1} + \sqrt Z  < \sqrt {{\beta _{up}}}  $}
      \State ${\phi _{n}^{}}=\eta $.
    \ElsIf {${d_1} >\sqrt Z+\sqrt {{\beta _{up}}}  $}
      \State ${\phi _{n}^{}}=(0,0)$.
     \ElsIf {${d_1} \ge \left| {\sqrt {{\beta _{up}}}  - \sqrt Z } \right|\& {d_1} \le \sqrt {{\beta _{up}}}  + \sqrt Z   $}
      \State If $\gamma$ satisfies $\overline {{\rm{C32}}}$, then ${\phi _{n}^{}}=\gamma$, else if $\eta$ satisfies
      \State $\overline {{\rm{C1}}}$, then ${\phi _{n}^{}}=\eta$, otherwise ${\phi _{n}^{}}$ is one of ${\varepsilon _1} $ and ${\varepsilon _2}$
      \State that makes $\overline {SINR_B}$ higher.
      \EndIf
  \EndIf
\end{algorithmic}
\end{algorithm}

\section{Simulation Results}
The effectiveness of the proposed active RIS aided surveillance scheme is verified by conducting numerical simulations. {\color{blue} Similar with \cite{9652031}, }we consider the Alan-Bob channel and the Alan-Eve channel are modeled as Rayleigh fading with path-loss exponent 3.5, the Alan-Rey channel, the Rey-Bob channel and the Rey-Eve channel are modeled as Rician fading with Rician factor 5 and path-loss exponent 2.2. The large-scale fading at reference distance of 1 m is -30 dB. The intervals between each reflecting element are half-wavelength. For the location of each node, we consider a two dimensional coordinate space, and let Alan, Bob, Rey, Eve to be fixed in the coordinate (0, 0), (8, 0), (7, 4), (5, 0) respectively. {\color{blue} Under the above coordinates, the average path losses of the  Alan-Bob channel, the Alan-Eve channel ,the Alan-Rey channel, the Rey-Bob channel and the Rey-Eve channel are -61.5 dB, -54.4 dB, -50 dB, -43.5 dB and -44.3 dB, respectively. $P_A$ and $P_{max}$ are normalized over the noises $\sigma _0^2 = \sigma _r^2 =1$.} The simulation results are collected from Monte Carlo simulations with 1000 independent channel realizations. { \color{blue}To serve as a benchmark, we give the eavesdropping rate of the passive RIS scheme under the same scenario, in which only the phases of the reflecting coefficient matrix are optimized. All initial points involved in the iteration are set to 0.01.}

Fig. 3 presents the eavesdropping rate as a function of $P_A$, in which $N_R=20$ and $P_{max}=60\rm\ dB$. As can be seen, the proposed scheme achieves much higher eavesdropping rate than the benchmark. That is because the active RIS can not only adjust the phase but also amplify the amplitude of the reflected signal, hence the reflected signal has enough power to adjust both $SINR_B$ and $SINR_E$. By contrast, in the benchmark, the RIS can only adjust the phase of the reflected signal, which means the signal $s_A$ experiences severe double path loss after passing through the Alan-Rey-Bob (Eve) link and compare with the signal passing through the directly Alan-Bob (Eve), the reflected signal becomes very weak. Also, although the suboptimal solution is lower than the optimal solution, it is still obviously batter than the benchmark and has lower computation complexity.

\begin{figure}[t]
  \centering
  \includegraphics[width=1.8in]{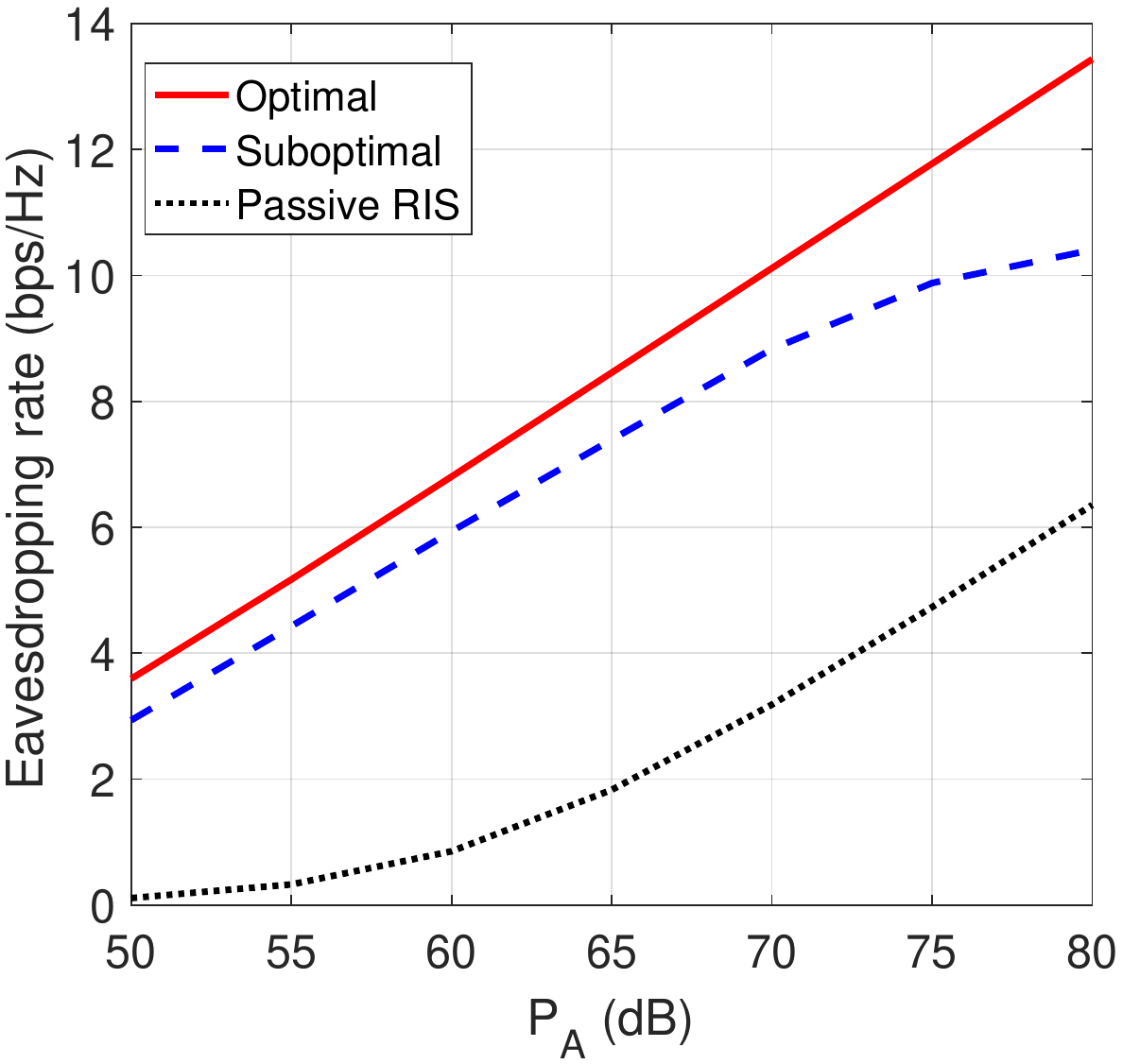}\\
  \caption{ The eavesdropping rate versus $P_A$, where $N_R=20$ and $P_{max}=60\rm\ dB$. }
\end{figure}
\begin{figure}[t]
  \centering
  \includegraphics[width=2.0in]{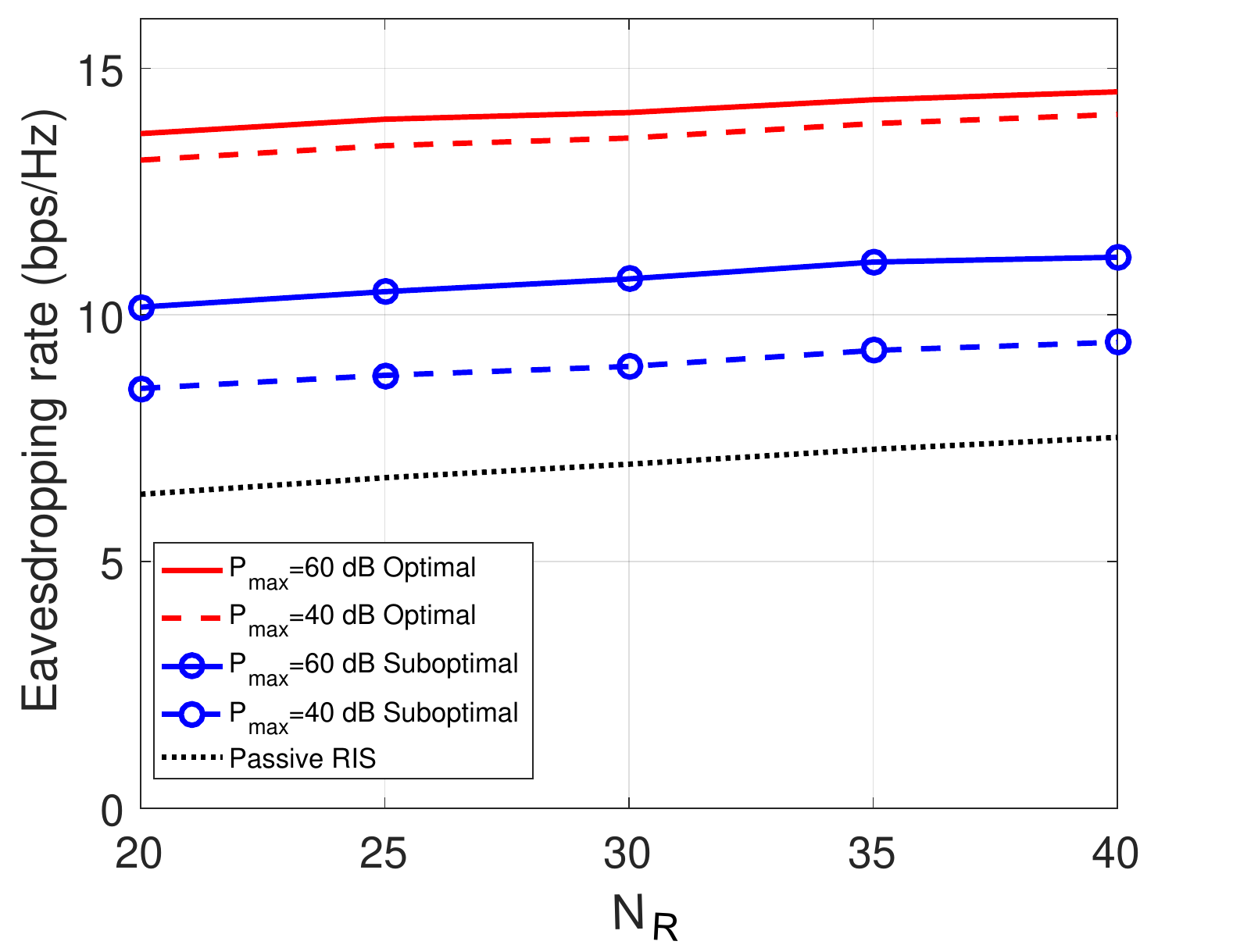}\\
  \caption{ The eavesdropping rate versus $N_R$ with different $P_{max}$, where $P_{A}=80\rm\ dB$.  }
\end{figure}

Fig. 4 represents the eavesdropping rate versus the number of the reflecting element of Rey with $P_A=80$ dB. Due to the increased degree of freedom, the eavesdropping rates of both proposed active RIS scheme and the passive RIS scheme increase with $N_R$. But the active RIS scheme is significantly batter than the passive RIS scheme, that is because the active RIS could effectively overcome the double path loss effect. As shown in Fig. 4, the eavesdropping rate increases with the maximum power budget of each element in Rey, $P_{max}$, that is because the higher $P_{max}$ means the higher received signal strength from Rey.

\section{Conclusion}
In this letter, an active RIS aided surveillance scheme was proposed for legitimate proactive eavesdropping. For maximizing the eavesdropping rate, we have optimized the reflecting coefficients of the active RIS. Then, we have proposed the corresponding
algorithm to solve the optimization problem by using a series of auxiliary variables. Simulation results have verified that our proposed active RIS aided surveillance scheme can effectively proactive eavesdrop the dubious communication link.


\bibliographystyle{IEEEtran}

\bibliography{ref}

\end{document}